# Space-time Engineering with *Lasetron* Pulses

M. Kozlowski[1], J. Marciak-Kozlowska[2]

[1] Institute of Experimental Physics and Science Teachers College of Warsaw University, Hoża 69, 00-681 Warsaw, Poland,
[2] Institute of Electron Technology, Al Lotników 32/46, 02-668 Warsaw, Poland

**Abstract**
The *LASETRON* project (*Phys. Rev. Lett.*, **88,** (2002), p. 074801-1) offers the possibility of the investigation of electron-positron structure of the space-time. Following our results (*From Quarks to Bulk Matter*, Hadronic Press, 2001) we analyze theoretically possibility of the penetration of zeptosecond laser pulses ($10^{-21}$ s) through space-time.

## 1. Introduction

In the paper [1] the theoretical project of the *LASETRON* was described. As was shown in paper [1] the $10^{-21}$ s (zeptosecond) laser pulses can be generated using petawatt lasers, while already available terawatt lasers may generate subattosecond laser pulses of $10^{-19}$ s. The pulses will be radiated by ultrarelativistic electrons driven by circularly polarized high-intensity laser fields. *LASETRON* pulses can be achieved by placing a solid particle or a piece of wire of subwavelength cross section in the focal plane of a superpowerful laser.

In the book [2] it was shown that the lifetime of the electron-positron pair is of the order of $10^{-17}$ s. Strictly speaking the lifetime of the order of the relaxation time $\tau$

$$\tau = \frac{\hbar}{m_e \alpha^2 c^2} \approx 10^{-17}\,\text{s}. \qquad (1)$$

In formula (1) $m_e = m_p$ is electron, positron mass, $\alpha$ is the fine structure constant and $c$ is the vacuum light velocity. As can be concluded from formula (1) the *zitterbegung* or *tremor* of space-time can be investigated with the *LASETRON* pulses, for the latter are shorter than the characteristic relaxation time.

For the time period $\Delta t < \tau$, i.e. for *LASETRON* pulse the vacuum of space-time is filled with the gas of electron-positron pairs with life-time $\approx 10^{-17}$ s. In that case the propagation of *LASETRON* pulse will be described by the Heaviside equation;

$$\frac{\partial^2 E}{\partial t^2} + \frac{\sigma}{\varepsilon_0}\frac{\partial E}{\partial t} = c_\gamma^2 \frac{\partial^2 E}{\partial x^2}. \qquad (2)$$

In equation (1) $\varepsilon_0$ – permittivity of free space-time, and $\sigma$ is the conductivity of free space-time, $c_\gamma$ is the photon velocity[*] in space-time and *E* is electric field (in one dimension).

In the seminal paper [4] F. Calogero described the cosmic origin of quantization. In paper [4] the *tremor* of the cosmic particles is the origin of the quantization and the characteristic acceleration of these particles $a \approx 10^{-10}$ m/s² was calculated. In our earlier paper [5] the same value of the acceleration was obtained and compared to the experimental value of the measured space-time acceleration [6]. In this paper we define the cosmic force — *Planck* force, $F_{\text{Planck}} = M_P a_{\text{Planck}}$ ($a_{\text{Planck}} \approx a$) and study the history of Planck force as the function of the age of the Universe.

---

[*] $c_\gamma$ is the photon speed in space-time filled with electron-positron pairs. The electron-positron fluctuations can change the $c$ to $c_\gamma$, i.e $c_\gamma \neq c$ [3].

Masses introduce a curvature in space-time, light and matter are forced to move according to space-time metric. Since all the matter is in motion, the geometry of space is constantly changing. A Einstein relates the curvature of space to the mass/energy density:

$$\mathbf{G} = k\mathbf{T}, \tag{3}$$

$\mathbf{G}$ is the Einstein curvature tensor and $\mathbf{T}$ the stress-energy tensor. The proportionality factor $k$ follows by comparison with Newton's theory of gravity: $k = G/c^4$ where $G$ is the Newton's gravity constant and $c$ is the vacuum velocity of light; it amounts to about $2.10^{-43} \text{N}^{-1}$, expressing the *rigidity* of space-time.

In paper [5] the model for the acceleration of space-time was developed. Prescribing the $-G$ for space-time and $+G$ for matter the acceleration of space-time was obtained:

$$a_{\text{Planck}} = -\frac{1}{2}\left(\frac{\pi}{4}\right)^{\frac{1}{2}} \frac{\left(N + \frac{3}{4}\right)^{\frac{1}{2}}}{M^{\frac{3}{2}}} A_P \tag{4}$$

where $A_P$, *Planck* acceleration equal, viz.:

$$A_P = \left(\frac{c^7}{\hbar G}\right)^{\frac{1}{2}} = \frac{c}{\tau_P} \cong 10^{51} \text{ms}^{-2}. \tag{5}$$

As was shown in paper [5] the $a_{\text{Planck}}$ for $N = M = 10^{60}$ is of the order of the acceleration detected by Pioneer spacecrafts [6].

Considering $A_P$ it is quite natural to define the *Planck* force $F_{\text{Planck}}$,

$$F_{\text{Planck}} = M_P A_P = \frac{c^4}{G} = k^{-1}, \tag{6}$$

where

$$M_P = \left(\frac{\hbar c}{G}\right)^{\frac{1}{2}}.$$

From formula (6) we conclude that $F_{\text{Planck}}^{-1}$ = rigidity of the space-time. The *Planck* force, $F_{\text{Planck}} = c^4/G = 1.2 \cdot 10^{44} N$ can be written in units which characterize the microspace-time, i.e. GeV and fm.
In that units

$$k^{-1} = F_{\text{Planck}} = 7.6 \cdot 10^{38} \text{GeV/fm}.$$

## 2. The *Planck*, *Yukawa* and *Bohr* forces

As was shown in paper [5] the present value of *Planck* force equal

$$F_{\text{Planck}}^{\text{Now}}(N = M = 10^{60}) \cong -\frac{1}{2}\left(\frac{\pi}{4}\right)^{\frac{1}{2}} 10^{-60} \frac{c^4}{G} = -10^{-22} \frac{\text{GeV}}{\text{fm}}. \quad (7)$$

In papers [7, 8] the *Planck* time $\tau_P$ was defined as the relaxation time for space-time

$$\tau_P = \frac{\hbar}{M_P c^2}. \quad (8)$$

Considering formulae (6) and (8) $F_{\text{Planck}}$ can be written as

$$F_{\text{Planck}} = \frac{M_P c}{\tau_P}, \quad (9)$$

where $c$ is the velocity for gravitation propagation. In papers [7, 8] the velocities and relaxation times for thermal energy propagation in atomic and nuclear matter were calculated:

$$\begin{aligned} v_{\text{atomic}} &= \alpha_{em} c, \\ v_{\text{nuclear}} &= \alpha_s c, \end{aligned} \quad (10)$$

where $\alpha_{em} = e^2/(\hbar c) = 1/137, \alpha_s = 0.15$. In the subsequent we define atomic and nuclear accelerations:

$$\begin{aligned} a_{\text{atomic}} &= \frac{\alpha_{em} c}{\tau_{\text{atomic}}}, \\ a_{\text{nuclear}} &= \frac{\alpha_s c}{\tau_{\text{nuclear}}}. \end{aligned} \quad (11)$$

Considering that $\tau_{\text{atomic}} = \hbar/(m_e \alpha_{em}^2 c^2)$, $\tau_{\text{nuclear}} = \hbar/(m_N \alpha_s^2 c^2)$ one obtains from formula (11)

$$\begin{aligned} a_{\text{atomic}} &= \frac{m_e c^3 \alpha_{em}^3}{\hbar}, \\ a_{\text{nuclear}} &= \frac{m_N c^3 \alpha_s^3}{\hbar}. \end{aligned} \quad (12)$$

We define, analogously to *Planck* force the new forces: $F_{\text{Bohr}}$, $F_{\text{Yukawa}}$

$$\begin{aligned} F_{\text{Bohr}} &= m_e a_{\text{atomic}} = \frac{(m_e c^2)^2}{\hbar c} \alpha_{em}^3 = 5 \cdot 10^{-13} \frac{\text{GeV}}{\text{fm}}, \\ F_{\text{Yukawa}} &= m_N a_{\text{nuclear}} = \frac{(m_N c^2)^2}{\hbar c} \alpha_s^3 = 1.6 \cdot 10^{-2} \frac{\text{GeV}}{\text{fm}}. \end{aligned} \quad (13)$$

Comparing formulae (7) and (13) we conclude that gradients of *Bohr* and *Yukawa* forces are much large than $F_{\text{Planck}}^{\text{Now}}$, i.e.:

$$\frac{F_{\text{Bohr}}}{F_{\text{Planck}}^{\text{Now}}} = \frac{5 \cdot 10^{-13}}{10^{-22}} \cong 10^9,$$

$$\frac{F_{\text{Yukawa}}}{F_{\text{Planck}}^{\text{Now}}} = \frac{10^{-2}}{10^{-22}} \cong 10^{20}. \quad (14)$$

The formulae (14) guarantee present day stability of matter on the nuclear and atomic levels.

As the time dependence of $F_{\text{Bohr}}$ and $F_{\text{Yukawa}}$ are not well established, in the subsequent we will assumed that $\alpha_s$ and $\alpha_{em}$ [9] do not dependent on time. Considering formulae (9) and (12) we obtain

$$\frac{F_{\text{Yukawa}}}{F_{\text{Planck}}} = \frac{1}{\left(\frac{\pi}{4}\right)^{\frac{1}{2}}} \frac{(m_N c^2)^2}{M_P c^2} \frac{\alpha_s^3}{\hbar} T, \quad (15)$$

$$\frac{F_{\text{Bohr}}}{F_{\text{Planck}}} = \frac{1}{\left(\frac{\pi}{4}\right)^{\frac{1}{2}}} \frac{(m_e c^2)^2}{M_P c^2} \frac{\alpha_{em}^3}{\hbar} T. \quad (16)$$

As can be realized from formulae (15), (16) in the past $F_{\text{Planck}} \approx F_{\text{Yukawa}}$ (for $T = 0.002$ s) and $F_{\text{Planck}} \approx F_{\text{Bohr}}$ (for $T \approx 10^8$ s), $T$ = age of universe. The calculated ages define the limits for instability of the nuclei and atoms.

## 3. The *Planck*, *Yukawa* and *Bohr* particles

In 1900 M. Planck [10] introduced the notion of the universal mass, later on called the *Planck* mass

$$M_P = \left(\frac{\hbar c}{G}\right)^{\frac{1}{2}}. \quad (17)$$

Considering the definition of the *Yukawa* force (13)

$$F_{\text{Yukawa}} = \frac{m_N v_N}{\tau_N} = \frac{m_N \alpha_{\text{strong}} c}{\tau_N}, \quad (18)$$

the formula (18) can be written as:

$$F_{\text{Yukawa}} = \frac{m_{\text{Yukawa}} c}{\tau_N}, \quad (19)$$

where

$$m_{\text{Yukawa}} = m_N \alpha_{\text{strong}} \cong 147 \frac{\text{MeV}}{c^2} \sim m_\pi. \qquad (20)$$

From the definition of the *Yukawa* force we deduced the mass of the particle which mediates the strong interaction – pion mass postulated by Yukawa in [11].

Accordingly for *Bohr* force:

$$F_{\text{Bohr}} = \frac{m_e v}{\tau_{\text{Bohr}}} = \frac{m_e \alpha_{em} c}{\tau_{\text{Bohr}}} = \frac{m_{\text{Bohr}} c}{\tau_{\text{Bohr}}}, \qquad (21)$$

$$m_{\text{Bohr}} = m_e \alpha_{em} = 3.7 \frac{\text{keV}}{c^2}. \qquad (22)$$

For the *Bohr* particle the range of interaction is

$$\gamma_{\text{Bohr}} = \frac{\hbar}{m_{\text{Bohr}} c} \approx 0.1 \, \text{nm}, \qquad (23)$$

which is of the order of atomic radius.

Considering the electromagnetic origin of the mass of the *Bohr* particle, the planned sources of hard electromagnetic field *LASETRON* [1] are best suited to the investigation of the properties of the *Bohr* particles.

## 4. Possible interpretation of $F_{\text{Planck}}$, $F_{\text{Yukawa}}$ and $F_{\text{Bohr}}$

In an important work, published already in 1951 J. Schwinger [12] demonstrated that in the background of a static uniform electric field, the QED space-time is unstable and decayed with spontaneous emission of $e^+ e^-$ pairs. In the paper [12] Schwinger calculated the critical field strengths $E_S$:

$$E_S = \frac{m_e^2 c^3}{e\hbar}. \qquad (24)$$

Considering formula (23) we define the *Schwinger* force:

$$F_{\text{Schwinger}}^e = eE_S = \frac{m_e^2 c^3}{\hbar}. \qquad (25)$$

Formula (24) can be written as:

$$F_{\text{Schwinger}}^e = \frac{m_e c}{\tau_{Sch}}, \qquad (26)$$

where

$$\tau_{Sch} = \frac{\hbar}{m_e c^2} \qquad (27)$$

is *Schwinger* relaxation time for the creation of $e^+ e^-$ pair. Considering formulae (13) the relation of $F_{\text{Yukawa}}$ and $F_{\text{Bohr}}$ to the *Schwinger* force can be established

$$F_{\text{Yukawa}} = \alpha_s^3 \left(\frac{m_N}{m_e}\right)^2 F_{\text{Schwinger}}^e, \quad \alpha_s = 0.15,$$

$$F_{\text{Bohr}} = \alpha_{em}^3 F_{\text{Schwinger}}^e, \quad \alpha_{em} = \frac{1}{137},$$
(28)

and for *Planck* force

$$F_{Planck} = \left(\frac{M_P}{m_e}\right)^2 F_{\text{Schwinger}}^e.$$
(29)

In Table 1 the values of the $F_{\text{Schwinger}}^e$, $F_{\text{Planck}}$, $F_{\text{Yukawa}}$ and $F_{\text{Bohr}}$ are presented, all in the same units GeV/fm. As in those units the forces span the range $10^{-13}$ to $10^{38}$ it is valuable to recalculate the *Yukawa* and *Bohr* forces in the units natural to nuclear and atomic level. In that case one obtains:

$$F_{Yukawa} \approx 16 \frac{\text{MeV}}{\text{fm}}.$$
(30)

It is quite interesting that $a_v \approx 16$ MeV is the volume part of the binding energy of the nuclei (droplet model).

Table 1: *Schwinger*, *Planck*, *Yukawa* and *Bohr* forces

| $F_{\text{Schwinger}}^e$ [GeV/fm] | $F_{\text{Planck}}$ [GeV/fm] | $F_{\text{Yukawa}}$ [GeV/fm] | $F_{\text{Bohr}}$ [GeV/fm] |
|---|---|---|---|
| $\approx 10^{-6}$ | $\approx 10^{38}$ | $\approx 10^{-2}$ | $\approx 10^{-13}$ |

For the *Bohr* force considering formula (13) one obtains:

$$F_{\text{Bohr}} \approx \frac{50 \text{eV}}{0.1 \text{nm}}.$$
(31)

Considering that the *Rydberg* energy $\approx 27$ eV and *Bohr* radius $\approx 0.1$ nm formula (31) can be written as

$$F_{\text{Bohr}} \approx \frac{\text{Rydberg energy}}{\text{Bohr radius}}.$$
(32)

## 5. Concluding remarks

In this paper the forces: *Planck*, *Yukawa* and *Bohr* were defined. It was shown that the present value of the *Planck* force (which is the source of the universe acceleration) $\approx 10^{-22}$ GeV/fm is much smaller than the *Yukawa* ($\approx 10^{-2}$ GeV/fm) and *Bohr* ($10^{-13}$ GeV/fm) forces respectively. This fact guarantees the stability of the matter in the present. However in the past for $T$ (age of the universe), $T < 0.002$ s, $F_{\text{Yukawa}} < F_{\text{Planck}}$ (0.002 s) and $F_{\text{Bohr}} < F_{\text{Planck}}$ ($10^8$ s). In this paper the relation of the *Schwinger* force (for the vacuum creation of the $e^+e^-$ pairs to the *Planck*, *Yukawa* and *Bohr* force was obtained. With the *LASETRON* pulses the electron-positron structure of space-time can be investigated.